\documentclass{llncs}
\usepackage[utf8]{inputenc}

\usepackage{graphicx}
\usepackage{xcolor}
\usepackage{verbatim}
\usepackage{svg}
\usepackage{url}
\usepackage{subcaption}
\usepackage{bbding}

\usepackage[ruled,vlined]{algorithm2e}

\makeatletter
\newcommand{\linebreakand}{%
  \end{@IEEEauthorhalign}
  \hfill\mbox{}\par
  \mbox{}\hfill\begin{@IEEEauthorhalign}
}
\makeatother

\title{Challenges and Opportunities for RISC-V Architectures towards Genomics-based Workloads}

\author{$\ast$Gonzalo Gómez-Sánchez\inst{2}\Envelope, $\ast$Aaron Call\inst{1}\Envelope,Xavier Teruel\inst{1},
Lorena Alonso\inst{1},
Ignasi Moran\inst{1},\\
Miguel Ángel Pérez\inst{1}, David Torrents\inst{1,3},
Josep Ll. Berral\inst{2,1}}
\institute{Barcelona Supercomputing Center (BSC)\\
\email{aaron.call, xavier.teruel, ignasi.moran, lorena.alonso, miguel.perez, david.torrents\}@bsc.es}
\and Universitat Polit\`{e}cnica de Catalunya - BarcelonaTECH\\
\email{\{gonzalo.gomez, josep.ll.berral\}@upc.edu}
\and Institut Català de Recerca i Estudis Avançats, Barcelona, Spain
}

\begin{document}
\Urlmuskip=0mu plus 1mu

\include{pythonlisting}

\maketitle


\begin{abstract}

%
The use of large-scale supercomputing architectures is a hard requirement for scientific computing Big-Data applications. An example is genomics analytics, where millions of data transformations and tests per patient need to be done to find relevant clinical indicators. Therefore, to ensure open and broad access to high-performance technologies, governments, and academia are pushing toward the introduction of novel computing architectures in large-scale scientific environments. This is the case of RISC-V, an open-source and royalty-free instruction-set architecture.
%
%
To evaluate such technologies, here we present the Variant-Interaction Analytics use case benchmarking suite and datasets. Through this use case, we search for possible genetic interactions using computational and  statistical methods, providing a representative case for heavy ETL (Extract, Transform, Load) data processing. Current implementations are implemented in x86-based supercomputers (e.g. MareNostrum-IV at the Barcelona Supercomputing Center (BSC)), and future steps propose RISC-V as part of the next MareNostrum generations.
%
%
Here we describe the Variant Interaction Use Case, highlighting the characteristics leveraging high-performance computing, indicating the caveats and challenges towards the next RISC-V developments and designs to come from a first comparison between x86 and RISC-V architectures on real Variant Interaction executions over real hardware implementations.

\end{abstract}



\section{Introduction}

\begin{figure}[h]
\centering
\captionsetup{font=footnotesize}
  \includegraphics[width=0.8\linewidth]{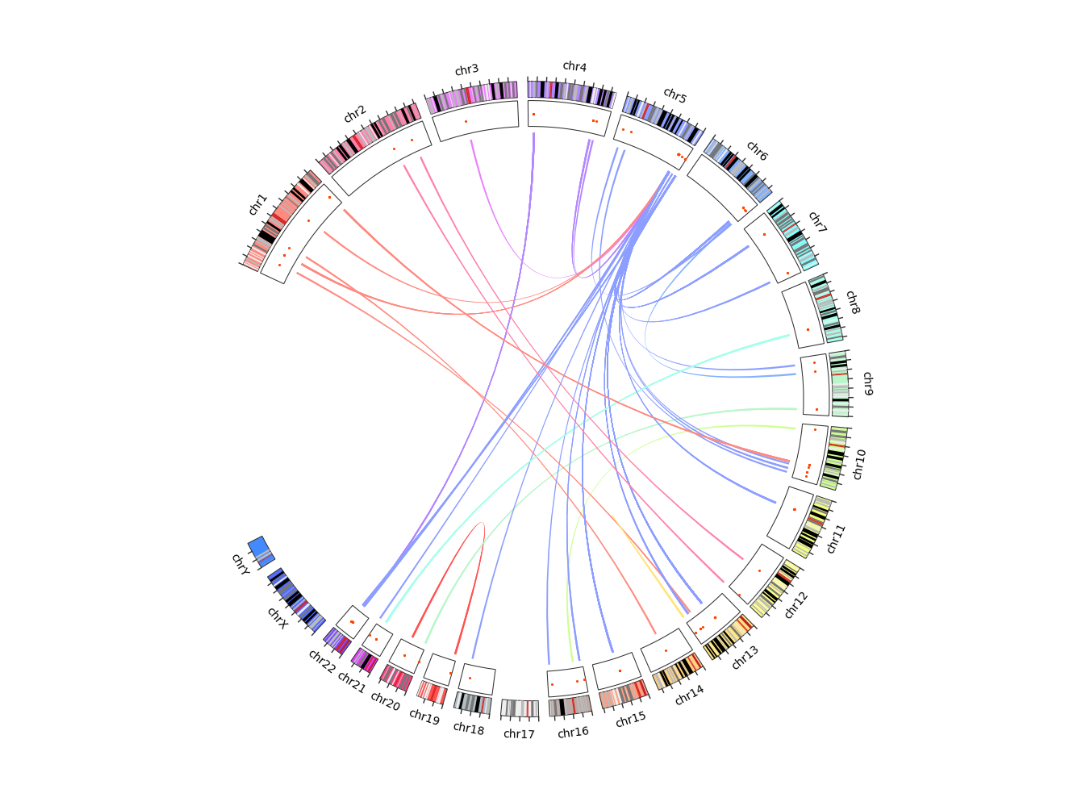}
  \caption{An example representation using CIRCOS \cite{circos} of different pairwise variant combinations.}
  \vspace{-2em}
  \label{fig:circos}
\end{figure}

Scientific applications depending on Big-Data processing rely on large-scale supercomputers to ingest large amounts of data towards the discovery of relevant clinical, environmental, social, or economics indicators at large and small scales (e.g. 1 PetaByte on genomics datasets~\cite{omics_petabytes}). This evidences the need for governments and research institutions' investment in supercomputing infrastructures to provide high-performance computing on efficiently managed resources. Current computing architectures are based on x86 and ARM, both proprietary and closed-source technologies. The limited access to such technologies for open research and hardware development generates a lack of trustworthiness regarding both privacy and security. To solve the mentioned issues, here we propose using RISC-V, an open-source and royalty-free instruction-set architecture.
During the last few years, there has been an increased interest in industry and academia to adopt RISC-V, mostly incentivized by strategic national/international benefits. As an example, the European Commission is pushing towards a mature RISC-V development, building an entire ecosystem around the European Processor Initiative~\cite{epi} (EPI), seeking technological sovereignty over the reduced group of non-European designers and manufacturers. The EPI project intends to design and manufacture the first entirely European chip based on an open standard, in collaboration with the local and international industry as well (i.e., Intel).
The software industry has already started to support RISC-V architectures, like Red Hat providing support for Linux Fedora distributions~\cite{fedora_riscv_arch,os_riscv} or Google announcing Android support for such~\cite{android_riscv}. Meanwhile, the hardware industry is moving towards adopting the RISC-V standard, collaborating with the design and manufacturing of newly RISC-V-based chips~\cite{press_release_rv_industry}. At the current time, functional prototypes are already on the market: for instance, HiFive~\cite{hifive} is a manufacturer of commercially available implementations of chip + board based on RISC-V, among others.\newline
RISC-V is becoming a natural alternative and a de facto standard for a new era of hardware, replacing current commodity chips. However, there is not yet a manufactured chip with equivalent characteristics to x86. Current RISC-V implementations lack fundamental extensions, and many applications lack support for specific hardware-depending features. Furthermore, other languages only support a subset of the RISC-V extensions, lacking vectorization or floating-point operations, for example. Therefore, benchmarking the proposed implementations requires new benchmarks focusing on the current state-of-art, which on one hand, are able to compare the existing capabilities and to point out the open challenges towards non-implemented ones, and on the other hand, are realistic towards their applicability in both academia and industry. The lack of standardization regarding how to benchmark workloads under this new architecture opens a new field of research in high-performance computing. As existing benchmarks on classic x86 architectures would not be fair for comparison against RISC-V in their state of implementation, here we propose a standard benchmark reaching the current capabilities of the architecture based on the real genomic analyses performed in supercomputing infrastructures for research. The goal is to be able to fairly compare the progress of the RISC-V development, identify the challenges and possibilities of improvement, and extract conclusions on the progress toward a complete and fully comparable RISC-V implementation.
In this paper, we present the Variant Interaction Analysis workload, proposed as a performance benchmark to explore the differences between the x86 and currently implemented RISC-V architecture, identifying the areas of improvement for the latter one regarding our scientific-based real HPC workload. We analyze the challenges RISC-V has to solve to perform as well as x86, also highlighting the principal \textit{to-do's}. Such analysis has not been done so far, and gathering this knowledge is crucial for the ongoing design of RISC-V prototypes as scientific HPC bound applications are one of the main targets that will benefit from such novel architectures.
The Variant-Interaction Analytics (VIA) workload, proposed here as a benchmark suite, is a genomics use case where a large number of pairwise combinations of genomic variants need to be analyzed to find its association with the disease~\cite{epistasis2}, one of the goals of human computational genomics.
Typically, the effect of each variant with the complex disease is studied one at a time. In variant interaction studies, the focus is on variants that interact with the complex disease depending on at least a second variant (see fig.\ref{fig:circos}). Methodologies such as Multifactor Dimensionality Reduction (MDR) combined with HPC technologies allow us to study the effect of pairwise combinations in large genomic datasets, helping us understand these diseases.
\vspace{-2em}
The number of combinations in this type of study can ascend to more than $10^{12}$. Due to the high number of tests to perform and the requirements in terms of resources consumption, performance, and volume of data (from storage to in-memory processing), we can consider that such an analytics application is HPC-oriented, along with a genomics background and a sensitive use case to analyze as a reference for Key performance Indicators (KPIs) evaluating real HPC workloads on RISC-V architectures. The main contributions of this presented work are:

\begin{itemize}
    \item Contribution 1: A Benchmark for Scientific HPC-based Analytics Application for RISC-V, adapted to the capabilities of current RISC-V implementations and designs, towards comparing with other established architectures and tracking its progress.
    \item Contribution 2: The identification of the challenges explaining the performance differences between RISC-V implementations and x86 on real HPC applications. 
    \item Contribution 3: A discussion and recommendations on the progress and improvement in RISC-V towards next step designs.
    \item Contribution 4: The creation of a publicly available open-data repository of benchmarks to run on RISC-V platforms~\cite{openrepo_rv_genomics}.
\end{itemize}

The rest of the paper is structured as follows: section~\ref{sec:related} introduces the state of the art, section~\ref{sec:methodology} discusses the standard datasets and benchmarking methodology, section~\ref{sec:evaluation} shows the performed evaluation. Finally, section~\ref{sec:discussion} concludes with a discussion and recommendations of challenges found and the next steps.

\section{Related Work}
\label{sec:related}

In the context of European sovereignty, the EU built an ecosystem around designing and manufacturing its chips. For this, they elected RISC-V as it is an open-source ISA. The cornerstone element is the European Processor Initiative (EPI) \cite{epi}. Around it, the MareNostrum Experimental Exascale Platform (MEEP) \cite{meep_paper} project proposes an open-source platform based on RISC-V to experiment on a RISC-V-based HPC ecosystem. MEEP is an open-source digital laboratory providing an ideal experimentation platform for RISC-V-based workloads. It integrates an accelerator that allows disaggregate computation from memory operations, optimizing the accelerator for dense (compute-bound) and sparse (memory-bound) workloads. MEEP invites software and hardware engineers to solve future challenges in the HPC, AI, ML, and DL domains. 

In the context of the "Designing RISC-V-based Acccelerators for next generation Computers" (DRAC)~\cite{predrac} project, authors on~\cite{bench_suite_riscv_gem5} implement a RISC-V vector instructions pipeline on the gem5 processor. Then they run a standard benchmark suite for HPC applications \cite{hpc_bench_suite} and adapt it to benchmark performance of vector instructions on RISC-V with a vector-length agnostic mechanism so that it is easily comparable with any other SIMD-compatible ISA (e.g., x86, ARM). 

RISER \cite{riser}, OpenCUBE \cite{oencube}, Vitamin-V \cite{vitamin-v}, and AERO \cite{aero} are four new EU-HORIZON projects that, combined together will provide a mature cloud environment for RISC-V. RISER will integrate low-power components and build an accelerator platform that includes the Arm-based Rhea processor from EPI and a PCIe acceleration board. OpenCUBE will provide a full-stack solution of a cloud computing blueprint deployed on European infrastructure, whereas Vitamin-V will deploy a complete hardware-software stack for cloud services based on cutting-edge and cloud open-source technologies for RISC-V and particularly focusing on EPI. Finally, the AERO project aims to bring and optimize the open-source software ecosystem to compile, runtimes, and auxiliary software deployment services on the cloud.  Vitamin-V will as well port a relevant benchmark suite for Big Data applications developed by the industry under the umbrella of the TPC council \cite{tpc} and their Big Data Analytics TPC-H benchmark.

Furthermore, RISC-V is also starting to be used for genomics. Wu et al.~\cite{dna-seq-riscv}, implement a RISC-V-based design on an FPGA that is later used for base-calling in DNA sequencing. DNA sequencing is the base method to obtain the genome sequence of an individual, thus making it fundamental for many bioinformatic applications, such as genomics. In this paper, they analyze using RISC-V as a better energy-efficient platform to sequence DNA. They achieve a 1.95x energy efficiency ratio compared to x86 architectures while being 38\% more energy-efficient than ARM architectures.

In the genomics field, the search for significant associations between genomic variants and complex diseases has been primarily studied using Genome-Wide Asociation Studies (GWAS)~\cite{gwas1}. These studies focus on the discovery of variants associated with the risk of developing the disease. This implies the use of a wide variety of methods including the logistic regression~\cite{logistic}, Bayesian partitioning~\cite{bayesian}, and other statistical methods~\cite{gwas3}.

While GWAS focuses on the detection of disease-susceptibility loci in a single independent manner, in variant interaction studies, the search is broadened to inspect the effects from the interaction of variants, which can be both additive or epistatic~\cite{epistasis1}. To tackle variant interactions, different methods, tools, and strategies have then arisen, focusing mainly on pairwise interactions~\cite{epistasismethods1}. Importantly, Multifactor Dimensionality Reduction (MDR)~\cite{multifactor1} is a supervised classification approach based on contingency tables that has become a reference in the field to reduce the dimension of the problem and to identify variant combinations associated with complex diseases.

\section{Benchmarking Methods}
\label{sec:methodology}

\subsection{Variant Interaction Analytics (VIA) Workload}
\begin{figure*}[ht]
\captionsetup{font=footnotesize}
  \includegraphics[width=\textwidth]{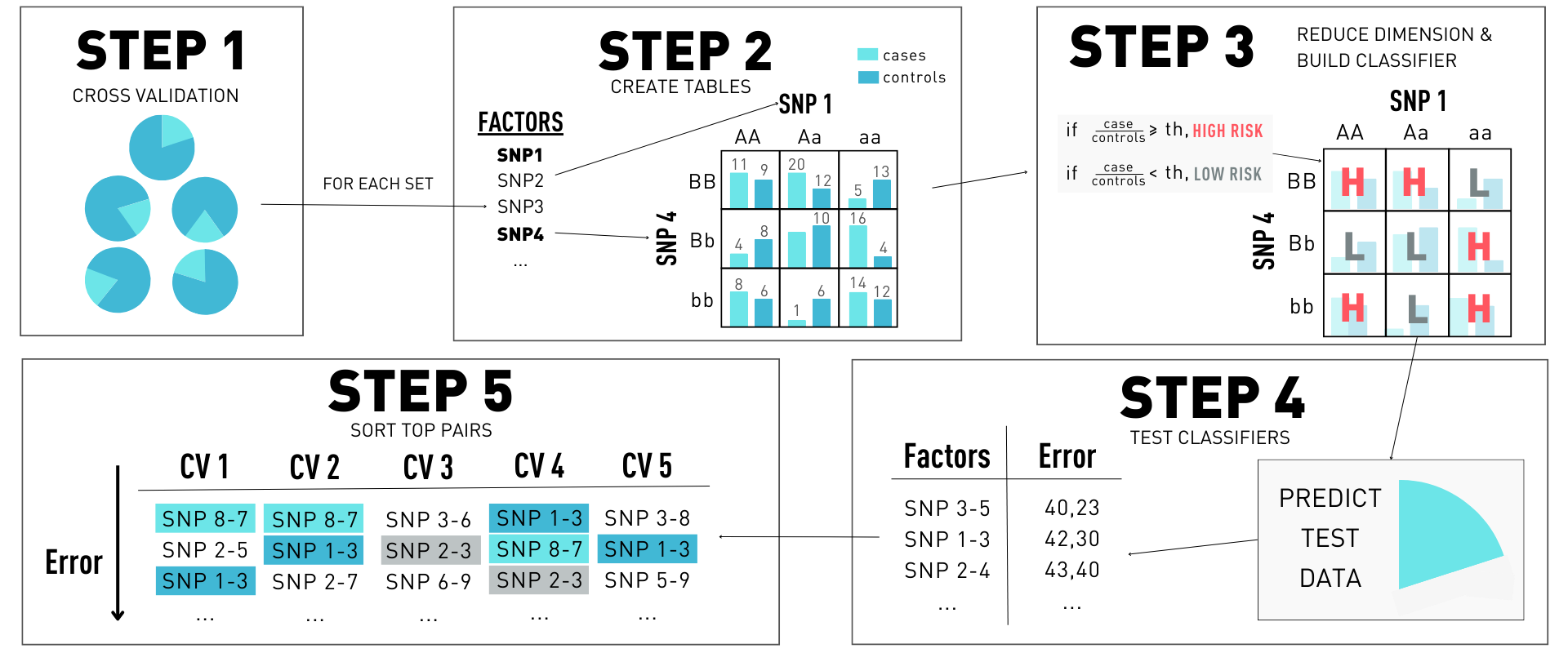}
  \caption{Multifactor Dimensionality Reduction algorithm steps. Step 1 is the cross-validation division. Step 2 is the building of contingency tables. In step 3, the dimension of the contingency tables is reduced to 1. In step 4, each multifactor is tested. Steps 2-4 are repeated for each cross-validation set and in Step 5, the top pairs are selected.}
  \vspace{-2em}
  \label{fig:mdr_steps}
\end{figure*}

The method we are applying in the Variant-Interaction Analytics (VIA) Workload is the MDR: a statistical method that, based on contingency tables, reduces the dimension of the problem. The variant-variant analysis, converts the counts obtained for the cases and controls into a simple binary variable by classifying all the possible allelic combinations for each pair in high-risk/low-risk, therefore, reducing the analysis to only one dimension. It follows a naive Bayes approach, building a probabilistic classifier from every variant-variant interaction and summarizing the best combinations for prediction. Figure~\ref{fig:mdr_steps} shows the 5 steps of the algorithm.
The process of computing a pair of variants does not imply a lot of memory or computational power. However, the high volume of pairs of variant combinations that we need to process demands the use of High-Performing Computing (HPC) technologies to make it feasible (see the description of the data at the end of this section). Since processing a pairwise combination of variants is computationally independent of the rest of the combinations, we can leverage the use of parallel computation frameworks. Furthermore, due to this independence of the combinations, we can also scale it down to test it in different architectures and extrapolate the results to real-case scenarios.

\subsection{Framework}

The method has been developed using Python and leveraging Apache Spark framework for parallel computation \cite{spark}. Apache Spark is an open-source distributed processing system for high-volume workloads. Thanks to its in-memory caching and a system of optimized queries, it can be used against data of any size. It has a master-slave architecture, combining a single master with multiple slaves.

\subsection{Dataset}

The dataset used for the benchmarking experiments is a synthetic dataset based on the Northwestern NuGENE project cohort \cite{nugene}. Maintaining the structure and the number of patients we have recreated a single part of a chromosome using randomized values for the new synthetic patients. The data is stored in compressed CSV format, following the structure in table \ref{tab:data}. 

\begin{table*}[]
\hspace{0em}
\centering
\captionsetup{font=footnotesize}
\begin{tabular}{ccccccccccccccccc}
\hline
chromosome & variant pos.   & Ref hom & Alt hom & AA & Aa & aa & AA & Aa & aa & ... & AA & Aa & aa \\
\hline

22         & 16231367   & A       & G       & 1       & 0       & 0       & 1       & 0       & 0       &     & 0       & 1       & 0  \\
22         & 17052123   & G       & A       & 0       & 1       & 0       & 0       & 0       & 1       &     & 0       & 0       & 1  \\
22         & 17055458   & G       & A       & 0       & 1       & 0       & 1       & 0       & 0       &     & 0       & 1       & 0  \\
\hline \\
\end{tabular}
\captionof{table}{First three rows of synthetic chromosome 22. The first four columns contain the identification of the variant while the rest of the columns contain the value of the variant per each patient of the 1,128 patients.} \label{tab:title} 
\vspace{-2em}
\label{tab:data}
\end{table*}

The labels of the data are saved in a different file, including the patient ID and a binary marker that indicates if the patient is a case or a control. A cohort such as NuGENE is composed of 11,297,253 variants, forming a dataset of 11,297,253 rows x 3,389 columns. In a variant interaction analysis, we are studying the association between each pairwise interaction with the disease, which means that in the most simple case, we are going to process every possible pair. This means that the number of combinations ascends approximately to (11,297,253 x 11,297,252)/2 = 63,813,957,024,378. The synthetic dataset created for the benchmark is a reduced version, generating 10 files of 50 SNPs with all the patients. We have not decreased the number of patients because maintaining the same structure allows us to easily extrapolate the computation time just by multiplying the number of combinations that need to be processed.

\vspace{-1.5em}
\section{Experiments}
\label{sec:evaluation}
\vspace{-1em}
\subsection{Evaluation Infrastructure}
\begin{figure}[htp]
\captionsetup{font=footnotesize}
    \subfloat[\scriptsize{Execution cycles}]{%
        \includegraphics[clip,width=0.5\columnwidth]{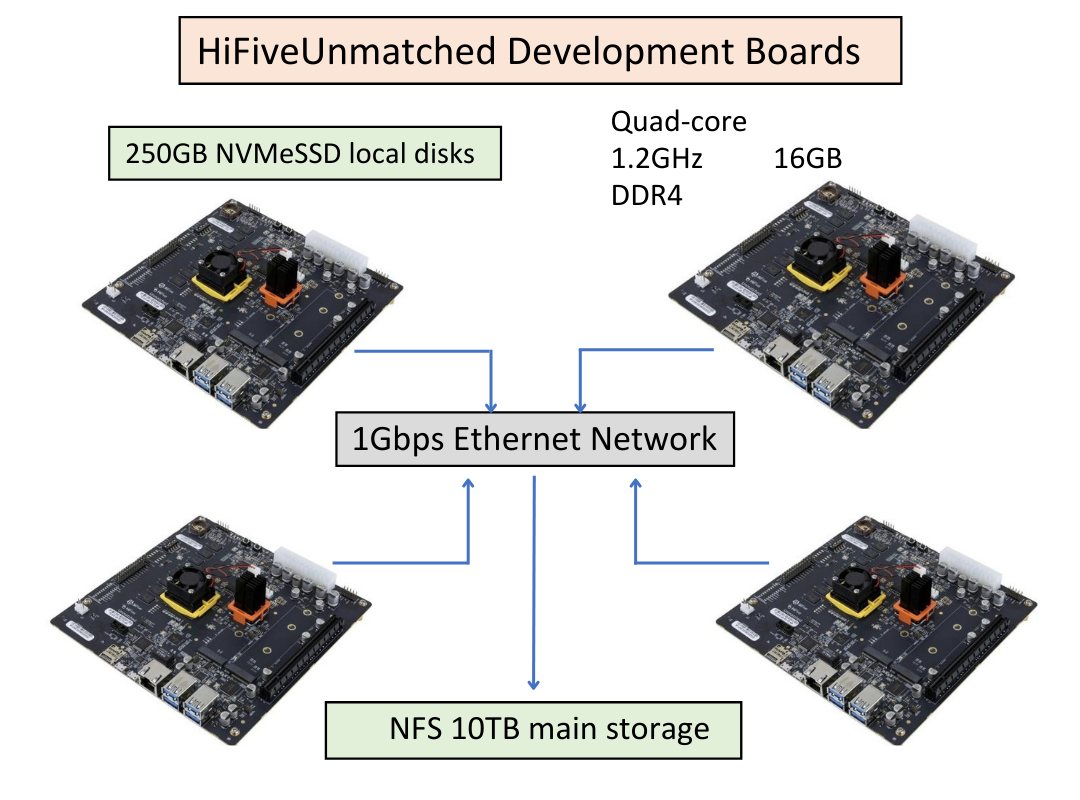}%
        \label{fig:unmatched-arch}
    }
    \subfloat[\scriptsize{Slow-down graph}]{%
        \includegraphics[clip,width=0.5\columnwidth]{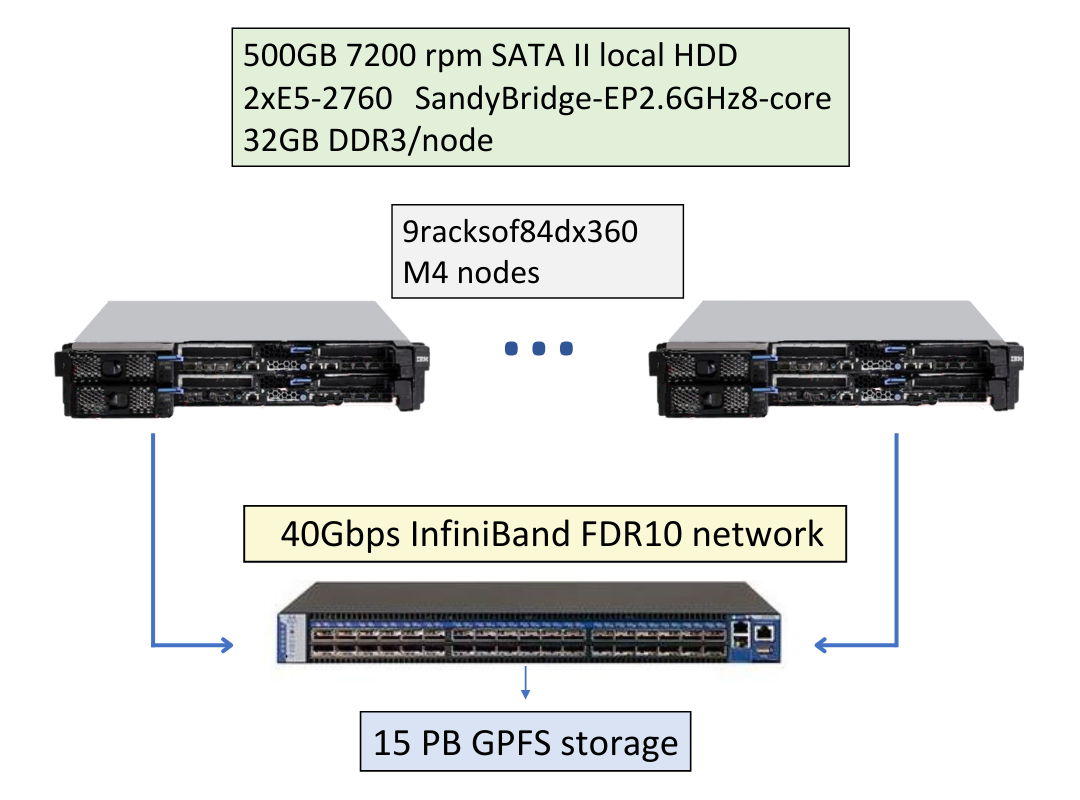}%
        \label{fig:nord3_openstack}
    }
\caption{Figure \ref{fig:unmatched-arch} shows Unmatched development architectures. While figure \ref{fig:nord3_openstack} shows Nord3 architecture, managed by OpenStack.}
\vspace{-2em}
\end{figure}

\vspace{-0.5em}
To perform the evaluation, we have deployed a cluster of up to four HiFive Unmatched development boards (figure \ref{fig:unmatched-arch}), comprising a quad-core RISC-V chipset operating at 1.2 GHz. The chip supports extensions IMAFDC, thus not including vectorial extensions. Each board has 16 GB of DDR4 and is interconnected with a 1Gbps ethernet network. 

On the other hand, our x86 cluster was composed of an OpenStack environment with four virtual machines with the following characteristics: 8 cores, 16GB of DDR3, and interconnected using an OpenStack Neutron network. Unlike the development boards, the actual hardware on the OpenStack is much more mature and is a production environment. The infrastructure (figure \ref{fig:nord3_openstack}) consists of SandyBridge-EP E5-2670, with eight cores at 2.6 GHz and 64GB DDR3 memory. The interconnection between nodes is comprised of an FDR10 InfiniBand network at 40Gbps. This has a relevant impact on understanding the performance divergences discussed in the next section. Notice that this is an outdated platform. However, we are comparing with a low-performance RISC-V infrastructure. Consequently, the comparison still allows us to understand the challenges that we need to solve to bring RISC-V to become commodity hardware.

We built a Java Zero VM 11 into RISC-V to support Apache Spark. When developing this work, no Java was fully ported to RISC-V. This was the reason for choosing Zero VM as the only feasible option. On the other hand, version 11 was chosen as it was the version some other workloads in the context of our project depended upon.
Moreover, support for Python runtime was enabled in RISC-V. This allowed us to run PySpark, a Python library to call Spark runtime (based on Java), on which the workload depends.
\vspace{-1.3em}
\subsection{Scalability Analysis}
To study the scalability of the workload in terms of data volume, we have performed experiments varying the Snumber of files. We have performed the expected 1, 3, and 5 files. Since we perform a pairwise combination of every file, each increment of files is more than a linear increment of the number of combinations processed, resulting in 2,500, 22,500, and 37,500 combinations respectively per case. 

To study the behavior in different environments, experiments have also been performed varying the number of nodes and cores available, simulating up to twelve different combinations. 

We scale per node and per core.
\begin{itemize}
    \item Per cores scalability: we scale up to four cores, as the RISC-V Unmatched boards only have up to four cores available.
    \item Per nodes scalability: we scale up to three nodes as worker nodes (under Spark point of view). That is because we had up to four RISC-V development boards. We could not test it using four nodes as workers because that would imply one node would need to share its resources with the master node. Doing so did not allow the experiments to run because of a lack of resources due to Spark runtime being a memory-intensive and compute-intensive runtime. It is designed for Big Data, but on RISC-V we have a more limited platform. Thus, we must adapt the execution tests to our platform's capabilities without sacrificing the process of the experiments.
\end{itemize}

Consequently, multiplying three combinations to scale nodes and up to four cores to scale production up to twelve simulations. Those simulations have been run on all the experiments described below (when relevant). 

We have compared the performance of both architectures in terms of time consumption using similar resources. Given the goal of the workload is to compute as many variant combinations as possible in the lesser time possible, execution time is an appropriate KPI. However, since the chipsets operating at each architecture uses different frequencies, execution time might give the false idea that x86 is always much better than RISC-V. Consequently, instead of comparing the execution time in the experiments, we have  normalized from seconds to cycles, achieving a fairer comparison. To do so, we have to multiply the execution time (seconds) by the frequency of each processor (Hertzs).

Regarding the dimension of the experiments, the selected number of files computed is up to five. This is because RISC-V takes an exponential time to compute a bigger amount of files. One could believe this is not realistic as, in a real-case scenario, the number of files will be much greater than that. However, the obtained metrics are valid and reasonable because each file is computed independently. Therefore, the times obtained are easily extrapolated to a real-case dataset.

The experiments performed are explained below.

    
\vspace{-1.3em}
\subsection{Vectorial vs Non-Vectorial}
\begin{figure}[h]
\centering
\captionsetup{font=footnotesize}
  \includegraphics[width=0.6\linewidth]{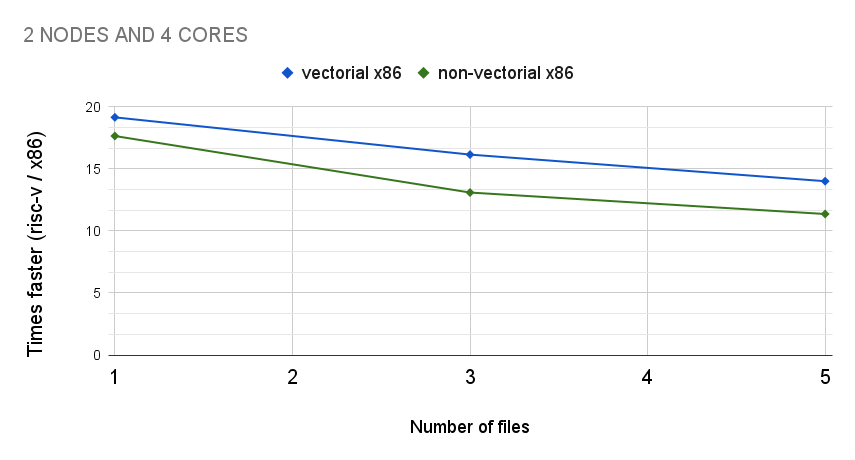}
  \caption{Slow-down graph of x86 using vector operations against RISC-V and x86 without vector operations against RISC-V. This example has been done using four cores and 2 workers, performing the runs for 1,3, and 5 files. }
  \vspace{-2em}
  \label{fig:vectorial_vs_nonvectorial}
\end{figure}

\vspace{-1em}
Initially, we run different experiments on both x86 and RISC-V platforms. 
In Figure~\ref{fig:vectorial_vs_nonvectorial} can be seen the comparison between the performance using two different versions for x86:
\begin{itemize}
    \item Vectorial: the original implementation for x86 architecture using vectorial operations leveraging numpy libraries. All the experiments performed in the 12 environments show that x86 performs at least more than 5 times faster than RISC-V.
    \item Non-vectorial: since RISC-V chipset is not supporting vectorial operations yet, we have tested and implemented a non-vectorial version of the workload for x86 to make a more fair comparison. The results show that the gap in cycles between both architectures gets smaller using this version. However, x86 is still at least three times faster than RISC-V.
    Most of the workload computational part relies on numpy to do the operations. Our non-vectorial x86 version disables it and attempts to use as many scalar operations as possible, achieving a suboptimal performance.
\end{itemize}

In both cases, the difference increase when the number of files is smaller. The reasons for this are, on one hand, because in RISC-V the time needed to load Spark and deploy the nodes is about 535 seconds (2.6 x 10\textsuperscript{10} cycles), 25 times slower in terms of cycles than in x86, which is less than 10 seconds (6.4 x 11\textsuperscript{11} cycles). On the other hand, we have disabled the vectorization on x86 via disabling numpy. However, this does not prevent the Python just-in-time (JIT) compiler from doing some optimizations of its own and using some vector instructions. Moreover, the Apache Spark runtime was not re-compiled without vector instructions, therefore, on the context initialization and some internal calls to it, there will be the presence of vector instructions as well. The reason for not to re-compile the Spark runtime is its complex build system, added to the complexity of disabling vector instructions on Java, having to do so both in the compiler and the JIT compiler. 

\subsection{Cores scalability}
\begin{figure}[h]
\captionsetup{font=footnotesize}
    \subfloat[\scriptsize{Execution cycles}]{%
        \includegraphics[clip,width=0.5\columnwidth]{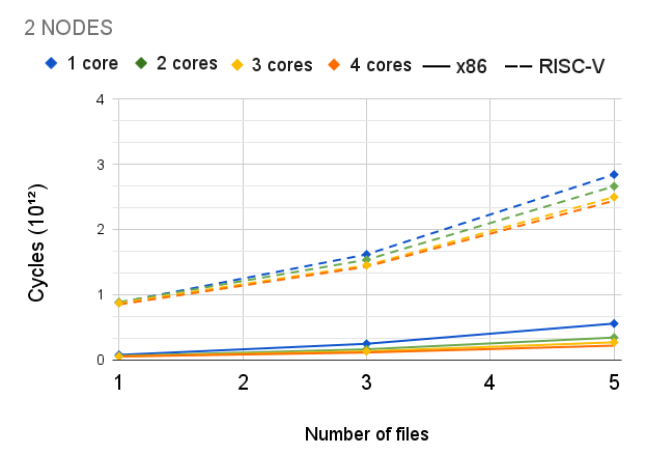}%
        \label{fig:cores_scalability_time}
    }
    \subfloat[\scriptsize{Slow-down graph}]{%
        \includegraphics[clip,width=0.5\columnwidth]{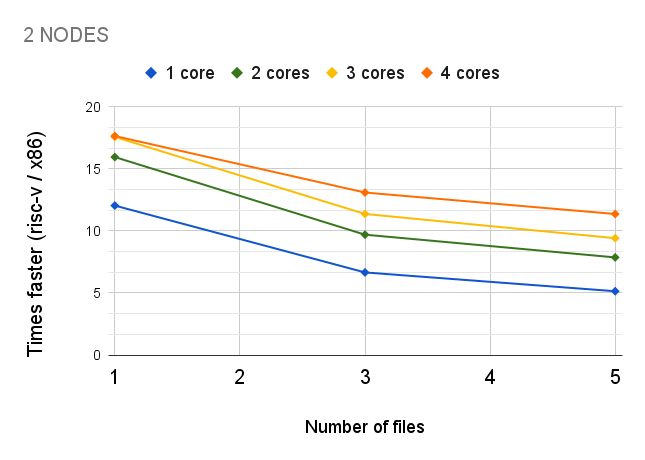}%
        \label{fig:cores_scalability_comp}
    }
\caption{Runtime (in cycles) comparison between non-vectorial x86 and RISC-V using different amounts of cores using 1,3 and 5 files.}
\vspace{-3em}
\end{figure}

 \vspace{-0.5em}
Based on our prior results, we set the non-vectorial x86 version as the fairer comparison version with our RISC-V boards. Looking into the performance as we increase the number of cores, it can be seen in Figure~\ref{fig:cores_scalability_time} how the cycles needed decrease as the cores do for both x86 and RISC-V. However, the gap between x86 and RISC-V is maintained and increased: x86 scales better in cores than RISC-V, as depicted in Figure \ref{fig:cores_scalability_comp}. The reasons for this gap increase are yet to be determined. To find out the reason one would need to examine the chip designs of both the x86 chipset and the RISC-V one. We know they are different, as one is a production chipset while the other is a development and rather immature. Their purposes are different. Hence would expect different capabilities in terms of cores-to-memory communication as well as cores-to-cache, and many other design elements. This exploration is left as future work, given this paper wants to focus more on the software side.  
\vspace{-2em}

\subsection{Nodes scalability}

\begin{figure}
\captionsetup{font=footnotesize}
    \subfloat[\scriptsize{Execution cycles in x86}]{%
        \includegraphics[clip,width=0.5\columnwidth]{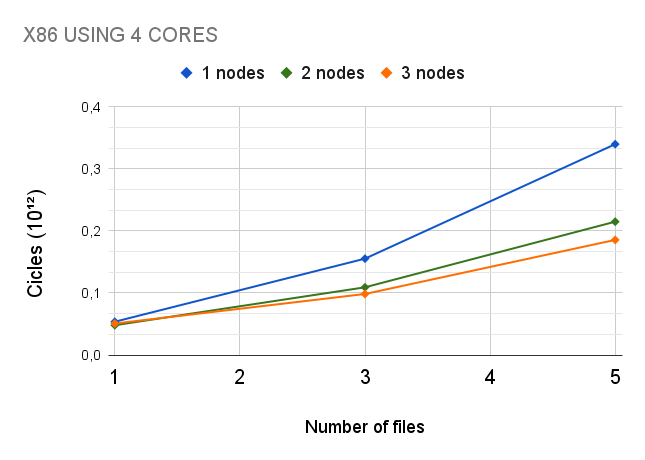}%
        \label{fig:workers_x86}
    }
    \subfloat[\scriptsize{Execution cycles in risc-v}]{%
        \includegraphics[clip,width=0.5\columnwidth]{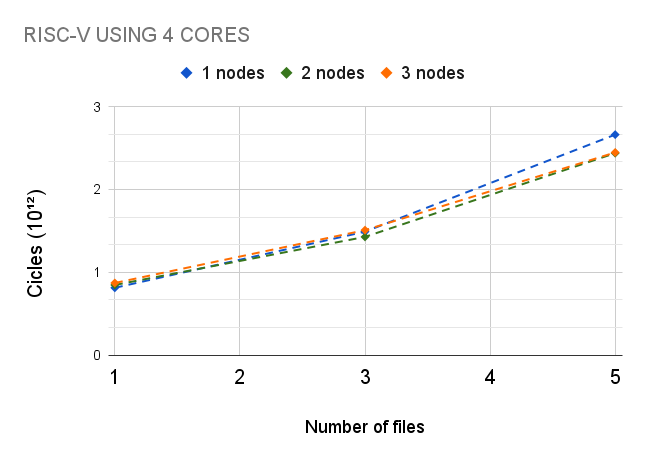}%
        \label{fig:workers_riscv}
    }
\caption{Execution cycles using a different number of nodes in x86 and RISC-V.}
\vspace{-2em}
\end{figure}

In figures \ref{fig:workers_x86} and \ref{fig:workers_riscv}, we show the scalability on the number of nodes (i.e, the number of worker nodes) on x86 and RISC-V platforms respectively. It is seen that when we increase the number of nodes, in x86 the number of cycles is decreased, scaling in nodes. This increase in performance is more notable as more files are processed. However, in RISC-V, we do not see any notable improvement in using more nodes. There is a slight decrease in cycles using more than one node processing five files, but no improvement at all when we go from 2 to 3. In fact, in most of the tests performed, going from two to three nodes produces a performance degradation on RISC-V.

\vspace{-2em}

\begin{figure}[h]
\centering
\captionsetup{font=footnotesize}
  \includegraphics[width=0.6\linewidth]{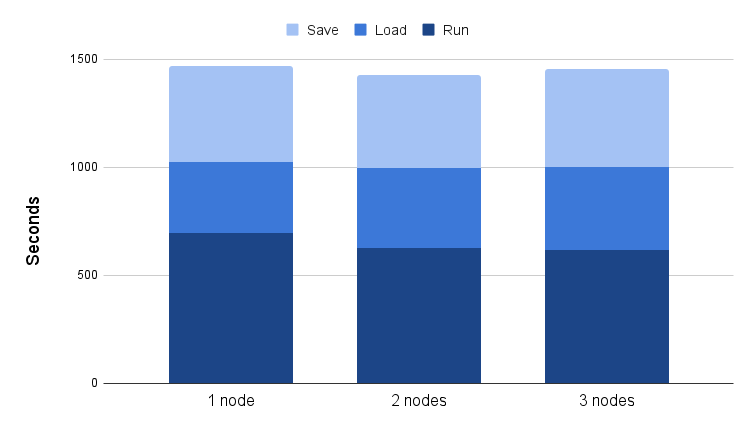}
  \caption{RISC-V Execution time for the different steps of the algorithm. For this experiment, 5 files were processed using 4 cores for each worker.}
  \vspace{-2em}
  \label{fig:runsaveload}
\end{figure}

To make a deeper analysis of what is happening in RISC-V, we have measured the execution time of the different steps of the algorithm. Figure \ref{fig:runsaveload} depicts the execution time of the different stages when running the VIA workload. While the run time of the workload is indeed slightly decreasing as we increase the number of nodes (i.e, workers), the total time is increasing due to the time expended in loading the data and saving the results.
One potential reason for this increased load time is that, in the current implementation, Spark produces the data splits and sends it to each of the worker nodes during execution time. Thus, prior to computing the real work, the node must have received the data previously. This split is done in the master node, so it does not matter whether or not we place the data into the local disks. In x86 we do have a production and powerful network interconnection, however, in RISC-V this network is more limited and not fitting so well for a Big Data use case. One solution for this would be placing a Hadoop Distributed File System (HDFS) below the runtime. This way, we would first load the data into HDFS, and at that point, the data would be split between the worker nodes. This would achieve the effect that when running the workload, the data would be already split and in place onto the worker nodes. So the time spent on the data loading stage would be minimal, and not have the impact we are currently experiencing. 
Finally, we have included all the results of the aforementioned experiments in an open-data repository, provisionally available at \cite{openrepo_rv_genomics}. In the near future, we plan to move it into a proper website with visualization of the different charts and metrics online. Moreover, the repository includes access to download the VIA workload as well as descriptions on how to reproduce it on your own platforms, either x86 or RISC-V. The repository and the workload are published under Apache License 2.0. 
\vspace{-1em}

\section{Recommendations and Discussion}
\label{sec:discussion}
\vspace{-1em}
From the conducted benchmarking experiments, RISC-V still presents an expected significant gap in terms of performance compared to x86 architectures. The principal challenge comes from the lack of vectorial extensions to be yet introduced or properly enabled on the RISC-V development boards. When disabling such extensions up to the feasible extent, the performance gap narrows. The remaining gap, below an order of magnitude, relies on the fact that available RISC-V platforms are in the research and development stages while x86 is already a trustworthy production environment. Thus, becoming the immediate next step to close the gap.
%
%
%

When benchmarking node scalability, we observe that data loading becomes a bottleneck for the tested RISC-V boards. Performed experiments show that RISC-V implementations scale properly as expected, but time improvement is eclipsed by the overhead of data loading. The proposed benchmark runs on Spark, which leverages in-memory data distribution, so when the blocking logic of data distribution is available, e.g., by using HDFS, worker nodes can quickly fetch data blocks concurrently. However, if such blocking logic is not available, e.g., reading a file in a regular File System, data must be fetched by the master node (first I/O bottleneck) and then shuffled across working nodes (second I/O bottleneck). The next step would be to port and enable HDFS or an equivalent format or DFS to allow us to observe the node scalability regardless of current differences in I/O performance between architectures. This includes as well network communication.
%

Summarizing, at this time the main challenge in comparing architectures is that there is no one-to-one match between instructions in x86 and the instructions available on currently implemented RISC-V chips. Moreover, other differences in design between chipsets challenge performing absolutely fair comparisons between platforms. For this reason, we are forced to use different metrics focusing more on metrics related to cycles/operation-per-analytics/data. 
%
%

The next steps towards refining the benchmarking process are, first, to isolate the data load time from the execution itself of the workload. And second, to have a compiled runtime and workload with equivalent instructions on both architectures, comparing the best implementation of the workload for both architectures.

\subsubsection*{Acknowledgements}
{\scriptsize
This work has been partially financed by the European Commission (EU-HORIZON  NEARDATA GA.101092644, VITAMIN-V GA.101093062), the MEEP Project whichreceived funding from the European High-Performance Computing Joint Undertaking (JU) under grant agreement No 946002. The JU receives support from the European Union's Horizon 2020 research and innovation program and Spain, Croatia and Turkey. Also by the Spanish Ministry of Science (MICINN) under scholarship BES-2017-081635, the Research State Agency (AEI) and European Regional Development Funds (ERDF/FEDER) under DALEST grant agreement PID2021-126248OB-I00, MCIN/AEI/10.13039/ 501100011033/FEDER and PID GA PID2019-107255GB-C21, and the Generalitat de Catalunya (AGAUR) under grant agreements 2021-SGR-00478, 2021-SGR-01626 and "FSE Invertint en el teu futur".
}



\bibliographystyle{splncs03}

\raggedright


\end{document}